# Theoretical study of spin-torque oscillator coupled with a nano-magnet by dipole-dipole interaction

H. Arai, T.Taniguchi, and H. Imamura

Spintronics R.C., National Institute of Advanced Industrial Science and Technology (AIST), Umezono 1-1-1,Tsukuba 305-8568, Japan

The dynamics of a spin-torque-oscillator (STO) coupled with a nano-magnet through dipole-dipole interaction was studied numerically by using the macrospin model for the application of the STO as a read head sensor of hard disk drives. We found that the current, which is required to induce the oscillation of the free-layer (FL) of the STO, depends strongly on the distance between the FL and the nano-magnet as well as on the relative orientation of the magnetizations between them. To determine the dynamics of the STO it is indispensable to consider the dynamics of the dipole-coupled nano-magnet. We showed that we could detect the orientation of the magnetization of a nano-magnet, or a recording-bit, by the modulation of the oscillation frequency of the STO.

## 1 Introduction

To keep up with the recent rapid progress of the information technology much effort has been devoted to increasing the recording density of the hard disc drives (HDDs). However since the size of a recording bit (RB) of conventional HDDs is reaching the limit[1], much attention has been focused on the three dimensional magnetic recording. A spin torque oscillator (STO) is one of the powerful candidates of the read head for the three-dimensional magnetic recording[2–5]. Braganca et al. [2] numerically showed that STO frequency is shifted by magnetostatic field from a recording bit. Mizushima et al. [3] and Kudo et al. [4] theoretically showed that STO senses the media field as a modulation in the oscillation frequency with high signal transfer rates. Suto et al. [5] experimentally showed that STO frequency is instantly shifted by applying a pulse magnetic field. These studies showed that the STO can detect a small magnetic field from the RB. However, it should be noted that in these studies they have implicitly assumed that the magnetization of the RB is fixed even if it couples with the free-layer (FL) of the STO through magnetic dipole-dipole interaction. Therefore, it is intriguing to study how the oscillation of FL is affected by the magnetization dynamics of the RB.

In this paper, we numerically studied the dynamics of the FL coupled with the RB through the dipole-dipole interaction based on the macrospin model. We found that the magnetization dynamics of the RB drastically changes the conditions of the current and distance between the FL and RB for the oscillation of FL. We showed that we could read the RB by the modulation of STO frequency.

## 2 Models and Methods

The system we considered is schematically shown in Fig.1. The FL and a RB are magnetically coupled with each other through dipole-dipole interaction. The spin polarized electric current was flowing in the positive $z$-direction exerts a spin torque on the FL.

In the macrospin model, the equations of motion of the FL and RB are given by the following Landau-Lifshitz-Gilbert (LLG) equation,

$$\frac{\partial m_1}{\partial t} = -\gamma \left( m_1 \times H_{\text{eff}}^{(1)} \right) + \alpha \left( m_1 \times \frac{\partial m_1}{\partial t} \right) - T_1, \quad (1)$$

$$\frac{\partial m_2}{\partial t} = -\gamma \left( m_2 \times H_{\text{eff}}^{(2)} \right) + \alpha \left( m_2 \times \frac{\partial m_2}{\partial t} \right), \quad (2)$$

where $\gamma$ is the gyromagnetic constant, $m_1$($m_2$) the unit-vector directing the magnetization of the FL (RB), α the Gilbert damping constant, and $T_1$ the Slonczewski spin torque[6]. Introducing the unit vector $p = (-1, 0, 0)$ directing the magnetization of the pinned layer (PL) the spin torque term is expressed as

$$T_1 = v\, m_1 \times (m_1 \times p), \quad (3)$$

where the spin torque coefficient v is defined as

$$v = \frac{g \mu_B I_s}{2 e M_s V}. \quad (4)$$

Here $g$ is the g-factor of electron spin, $\mu_B$ the Bohr magneton, $I_s$ the spin polarized current, $e$ the absolute value of the electron charge, $M_s$ the saturation magnetization of the FL, $V$ the volume of the FL. The effective field $H^{(i)}_{\text{eff}}$ is given by the sum of the anisotropy field, the demagnetization field, and the dipole field as

$$H_{\text{eff}}^{(i)} = H_{\text{anis}}^{(i)} + H_{\text{demag}}^{(i)} + H_{\text{dipole}}^{(j \to i)}. \quad (5)$$

For simplicity, we assumed that both the FL and RB were the same shape (thin elliptic disk) and made of the same ferromagnetic material. We assumed that the anisotropy field was given by $H^{(i)}_{\text{anis}} = (H_k m^{(i)}_x, 0, 0)$, and the demagnetization field was given by $H^{(i)}_{\text{demag}} = (0, 0, -M_s m^{(i)}_z)$. The dipole field is expressed as

$$H_{\text{dipole}}^{(j \to i)} = -\frac{\mu_j}{4\pi r^3}[m_j - 3(m_j \cdot \hat{r})\hat{r}], \quad (6)$$

where $\mu_j$ is the magnitude of the magnetic moment of j ($\neq$i), $r$ is the distance between $m_i$ and $m_j$, $\hat{r}$ is the unit vector directing from $m_j$ to $m_i$. We set $\hat{r}$=(0, 0, -1) for $H^{(2 \to 1)}{}_{\text{dipole}}$, or $\hat{r}$=(0, 0, 1) for $H^{(1 \to 2)}{}_{\text{dipole}}$.

The positive direction of the current was taken to be from top to bottom in STO; i.e., electrons flow from the PL to the FL as shown in Fig. 1. We used the following material parameters of Py for the FL and RB [7]: $M_s$ = 0.8×10$^6$ [A/m]; $H_k$ = 3.0×10$^4$[A/m]; $\alpha$ = 0.01. We assumed a thin elliptic disk with a long radius of 20 nm, a short radius of 10 nm and a thickness of 2 nm; i.e., volume $V$ = π ×20×10×2 [nm³]. The oscillation frequency of the FL was determined from the peak of the power spectrum of $m^{(1)}{}_x$ (t).

## 3 Results and discussion

Let us first show the results obtained under assumption that the magnetization of the RB is fixed. Thus Eq. (2) is zero and the effect of $m_2$ on Eq. (1) is taken into account by the constant dipole field $H^{(2 \to 1)}{}_{\text{dipole}}$. We considered two initial configurations. One is the anti-parallel (AP) configuration where the magnetizations of the FL and RB are aligned to be AP to each other as shown in the inset of Fig. 2 (a). The other is the parallel (P) configuration shown in the inset of Fig. 2(b). In both configuration, we set p = (-1,0,0), and the magnetization direction of FL is slightly tilted from x-axis. We calculated the dynamics of the FL as a function of the polarized current, $I_s$, and the distance between the FL and RB, $r$, based on Eqs. (1) and (2).

Figure 2 (a) shows the results for the AP configuration. The dynamics of the STO can be classified into the following four regions: no STO dynamics, the in-plane oscillation of the FL, out-of-plane oscillation of the FL, and the switching of the FL. The threshold currents for the oscillation (switching) of the FL are plotted by the filled red (blue) circles against the distance between FL and RB. The open red circles represent the threshold currents for the out-of-plane oscillation of the FL. Dotted lines are guide for eyes. Following Ref. [7] the threshold current below that we cannot excite any STO dynamics was obtained as

$$I_{AP}^{fixed} = \frac{2e\mu_1\gamma}{g\mu_B}\alpha\left(\frac{1}{2}M_s + H_k + \frac{\mu_2}{4\pi r^3}\right), \qquad (7)$$

where $\mu_{1(2)}$ is the magnetic moment of the FL (RB). The last term of Eq. (7) is the contribution of the dipole coupling field from the RB. Red solid line shows the threshold current $I^{fixed}_{AP}$ of Eq. (7), which agrees well with numerical results indicated by red circles. Below the red line; i.e., in the white region, no STO dynamics was observed. In the red and yellow regions the in-plane and out-of-plane oscillations of the FL were induced by the spin torque, respectively. In the blue region the FL was switched to be parallel to the PL. In the limit of r→∞, the threshold currents of the STO oscillation and the switching of the FL are 16.5 µA and 19.7 µA, respectively, which agree with those of the STO without the RB[7].

The results for the P configuration are shown in Fig. 2 (b). Since the dipole field has the opposite sign to that of the AP configuration the threshold currents decrease with decreasing $r$. The analytic formula of the threshold current of the STO below which we can not excite the STO dynamics was obtained as

$$I_{\text{P}}^{\text{fixed}} = \frac{2e\mu_1\gamma}{g\mu_{\text{B}}}\alpha\left(\frac{1}{2}M_s + H_k - \frac{\mu_2}{4\pi r^3}\right), \qquad (8)$$

which also agrees with the numerical results as shown in Fig. 2 (b). One may expect that Figs. 2 (a) and (b) will give us the conditions for using the STO as a read head. However, as we shall show below, the oscillating regions shown in 2 (a) and (b) were drastically changed by considering the effect of the dynamics of the RB.

Next, we analyzed the STO dynamics with taking account of the dynamics of the RB by numerically solving the coupled differential equations (1) and (2).

Figure 3(a) shows the results for the AP configuration. We obtained the white, blue and red region similar to Fig.2 (b). However the positions of these regions in the $r$-$I_s$ plane were drastically changed. We found two oscillating regions: one is from $r = 6$ nm to 33 nm, the other is from $r =43$ nm to 90nm. From $r =34$nm to 42nm, we have no STO oscillation and the FL was switched to be parallel to the PL once the spin polarized current exceeds the threshold value. Both threshold currents for oscillation and switching increase with decreasing $r$ because the STO should overcome the damping of the STO and RB. From $r = 6$ nm to 33 nm the FL and RB are strongly coupled and therefore the damping torque felt by STO becomes twice as large as that without the RB. Therefore the threshold current for oscillation also becomes about twice as large as that given by Eq. (7).

Figure 3(b) shows the results for the P configuration. From $r = 7$ nm to 14 nm, we have the brown region where the RB was switched to the opposite direction to its initial state; i. e., the RB can not maintain its information during the reading process.

In order to see the difference between the oscillating regions of the AP and P configurations, Figs. 3 (a) and (b) were superimposed in Fig. 4 (a). In the green region, the STO oscillation occurred only for the AP configuration. In the yellow region, the STO oscillation occurred for both the AP and P configurations. In the purple region, the STO oscillation occurred only for the P configuration. We should exclude the shaded area corresponding to the brown region of Fig. 3(b) from the following discussion because we are interested in the ability of the STO as a read head. The results shown in Fig. 4(a) imply that we can read the RB by the modulation of the oscillation frequency of the STO if $I_s$ and $r$ are in the green, yellow, or purple regions.

Let us consider the reading process of the sequence of recording bits shown in Fig. 4(b) and the corresponding oscillation frequencies of the STO. If $I_s$ and $r$ are in the green region the STO oscillates only for the AP configuration; i. e., the STO frequency would be 0 for the P configuration and $f_{AP}$ for the AP configuration as shown in Fig. 4(c). On the other hand, if $I_s$ and $r$ are in the purple region the STO oscillates only for the P configuration with $f_P$ and the STO frequency varies as shown in Fig. 4 (d). In these green (purple) regions, the oscillation stops once the STO reads the RB in the P (AP) configuration. Thus we need an extra time for the STO to reach a stable oscillation.

In the yellow region, however, the STO oscillates for both the AP and P configurations with different oscillation frequencies of $f_{AP}$ and $f_P$, and the frequency of the STO varies as shown in Fig. 4 (e). Since the STO do not stop during the reading process this yellow region is suitable for high-speed reading. For example if we set $r$ = 30 nm and $I_s$ = 34μA, the STO frequencies are $f_{AP}$ = 10.9 GHz and $f_P$ = 9.1GHz, and the frequency shift is $\Delta f$ = 1.8 GHz which is large enough to read the RB.

## 4 Summary

We studied the dynamics of STO coupled with the RB through dipole-dipole interaction based on the macrospin model. We found that the dynamics of STO is strongly affected by the dynamics of the RB. We showed that there exist the values of $r$ and $I_s$ at which the STO act as a high-speed reading head.

**Acknowledgements**  We would like to thank H. Kubota, K. Ando, H. Saito, R. Matsumoto, H. Maehara, A. Emura, Y. Suzuki, R. Sato, K. Kudo, T. Yang, and H. Suto for valuable discussions. This work was partially supported by Strategic Promotion of Innovation Research and Development from Japan Science and Technology Agency (JST) and by JPSP KAKENHI Grant-in-Aid for Scientific Research (S) 23226001.

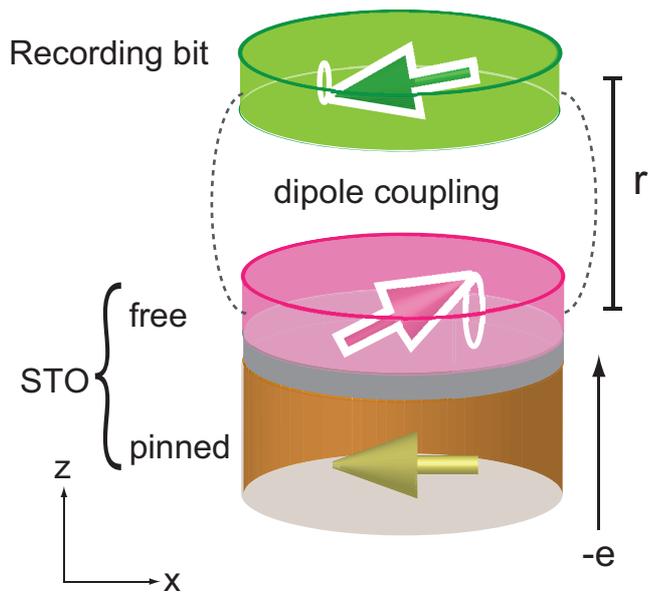

Fig.1

A schematic illustration of a STO read head and a RB. They are coupled with each other through magnetic dipole-dipole interaction. The distance between the FL and RB is $r$. Electrons flow from the pinned layer to the FL in STO to exert the spin torque on the FL. Then the STO excites the oscillation of the RB through the dipole coupling.

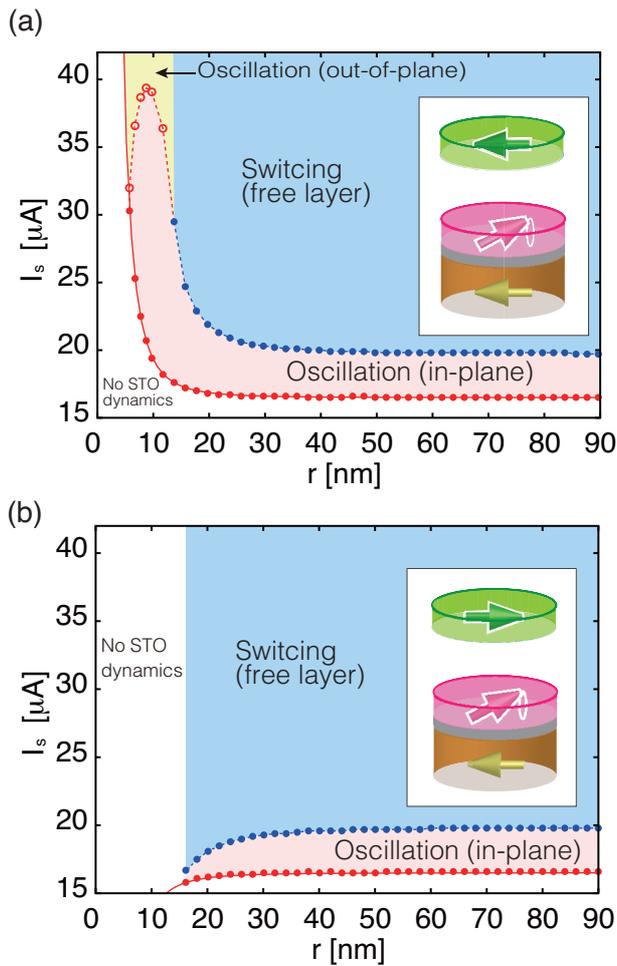

Fig.2

(a) A map of the dynamics of the FL against the distance between the FL and RB, $r$, and applied spin polarized current, $I_s$. The initial state was assumed to be the AP configuration as shown in the inset and the magnetization of the RB was fixed. Red and blue filled circles indicate the calculated threshold currents for oscillation and switching of the FL. Red open circles indicate the calculated threshold currents for out-of-plane oscillation of the FL. The red line represents the analytically obtained threshold currents of the FL dynamics, $I^{\text{fixed}}_{\text{AP}}$. The dotted lines are guide for eyes. In the white region the FL did not move; i.e., no STO dynamics were observed. In the red region, the FL oscillates around the $x$-axis (in-plane). In the yellow region, the FL oscillates around the

$z$-axis (out-of-plane). In the blue region the FL is switched to be parallel to the PL. (b) The same plot for the P configuration.

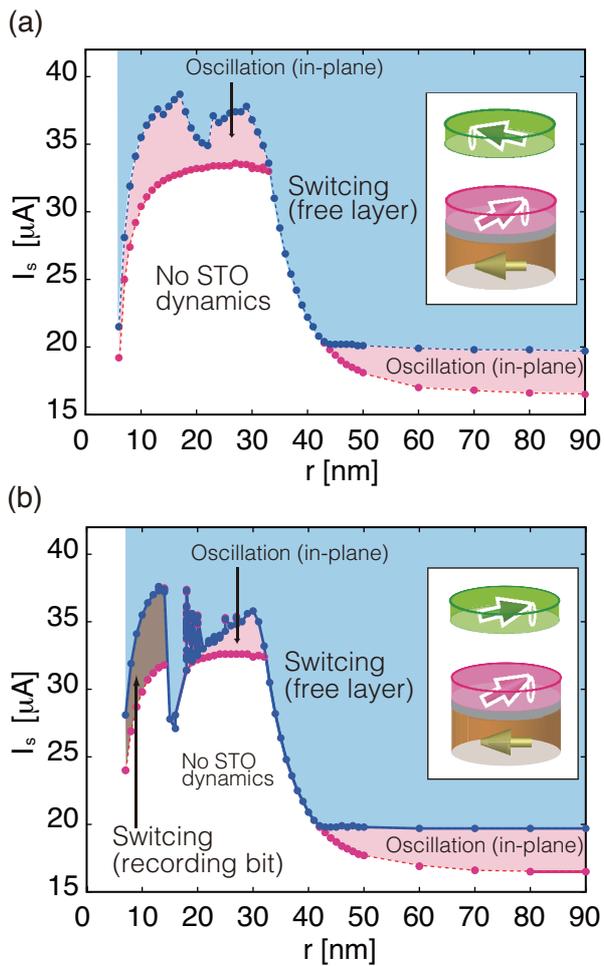

Fig.3

(a) A map of the dynamics of the FL against the distance between the FL and RB, $r$, and applied spin polarized current, $I_s$. The initial state was assumed to be AP configuration as shown in the inset. Red and blue filled circles indicate the calculated threshold currents for oscillation and switching of the FL. The dotted lines are guide for eyes. In the white region the FL did not move; i.e., no STO dynamics were observed. In the red region, the FL oscillates around the $x$-axis. In the blue region the FL is switched to be parallel to the PL. (b) The same plot for the P configuration.

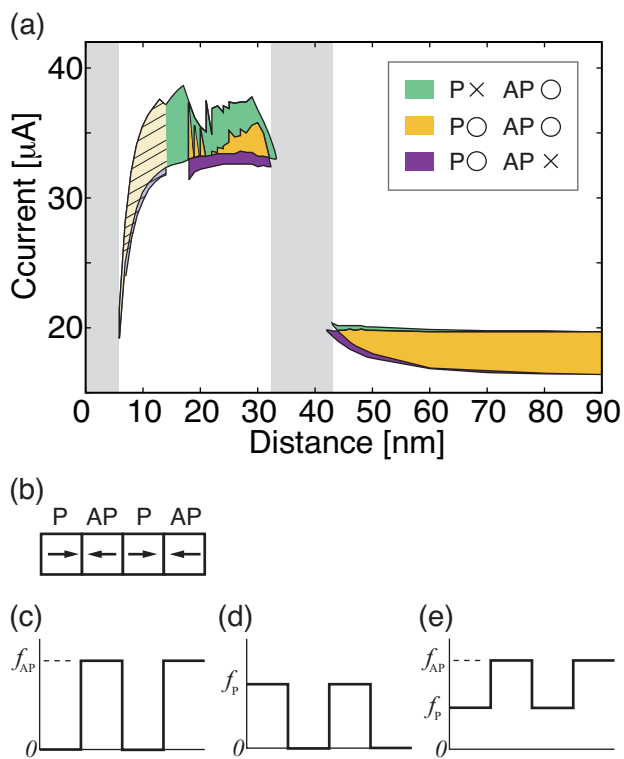

Fig.4

(a) Oscillating conditions of the STO. In the green (purple) region, the STO oscillates only for the AP (P) configuration. In the yellow region, the STO oscillates for both the AP and P configurations. In the gray region the STO has no oscillation. Shaded region corresponds to the brown region in Fig.3. (b) where the RB was switched to the opposite direction to its initial state. (b) A sequence of RBs. (c) The frequency of the STO in the green region while reading the sequence of RBs shown in (b). (d) The same plot in the purple region. (e) The same in the yellow region.